\documentclass{aa}  

\usepackage{graphicx}
\usepackage[colorlinks,urlcolor=blue,citecolor=blue,linkcolor=blue]{hyperref}
\usepackage{adjustbox}
\usepackage{subfig}
\usepackage{siunitx}
\usepackage{hyperref}
\usepackage{txfonts}
\usepackage{xcolor}

\def\R23{\mbox{$\rm R_{23}$}}

\def\arcsec{\hbox{$^{\prime\prime}$}}

\def\msun{M$_{_{\odot}}$}

\DeclareSIUnit{\mas}{mas}
\DeclareSIUnit{\dex}{dex}

\defcitealias{paperI}{Paper~I}

\usepackage{ulem}

\begin{document}

\title{MUSE-DARK}
\subtitle{~III: The evolution of the radial acceleration relation  at intermediate redshifts}
\titlerunning{RAR at intermediate-$z$}

\author{B. I. Ciocan
          \inst{1}
	\and 
	N. F.  Bouché
	\inst{1}
	\and
        J. Fensch
             \inst{1}
          D. Krajnović
         \inst{2}
         \and
        J. Freundlich
          \inst{3}
          \and
         H. Desmond
          \inst{4}
          \and 
          B. Famaey
          \inst{3} 
          \and
          R. Techi
          \inst{3}}
             \institute{Université Lyon 1, ENS de Lyon, CNRS, CRAL, UMR 5574, Saint-Genis-Laval, France\\
    \email{bianca-iulia.ciocan@univ-lyon1.fr} 
    \and 
    Leibniz-Institut für Astrophysik Potsdam (AIP), An der Sternwarte 16, 14482 Potsdam, Germany
  \and
 Univ. de Strasbourg, CNRS, Observatoire astronomique de Strasbourg, 11 rue de l’Université, 67000, Strasbourg, France
    \and
    Institute of Cosmology \& Gravitation, University of Portsmouth, Dennis Sciama Building, Portsmouth, PO1 3FX, UK
    }
   \date{}

\abstract{
\textit{Context.} The radial acceleration relation (RAR) is a tight empirical correlation between the observed radial acceleration ($a_{\rm{tot}}$) and the baryonic radial acceleration ($a_{\rm{bar}}$) measured across galaxy radii: these two accelerations start to deviate significantly from each other below a characteristic acceleration scale, $a_0$.
So far, observational studies of the RAR have predominantly focused on galaxies in the local Universe, leaving its evolution with cosmic time largely unexplored. \\  
\textit{Aims.}  Using high signal-to-noise data from the MUSE Hubble Ultra Deep Field survey, 
we investigate the RAR with a sample of 79 star-forming galaxies (complete above $M_\star>10^{8.8}$~\msun) at intermediate redshifts ($0.33 < z < 1.44$).\\  
\textit{Methods.} We estimate the observed intrinsic acceleration ($a_{\rm{tot}}$) and 
the baryonic acceleration ($a_{\rm{bar}}$) from a disk-halo decomposition that incorporates stellar, gas, and dark matter components, with corrections for pressure support,  using 3D forward modelling.\\  
\textit{Results.} We find a RAR in our intermediate-$z$ sample offset from the local relation, with a higher characteristic acceleration scale ($a_0|_{z\sim1} = 2.38 ^{+0.12}_{-0.10} \times 10^{-10}~\mathrm{m/s^2}$) and a larger intrinsic scatter ($\sim0.17$ dex). Dividing the sample into redshift bins and refitting the RAR in each bin, we find a characteristic acceleration scale that systematically increases with $z$. Parametrizing the $z$-dependence as $a_0(z) = a_{0}(0) + a_{1}\cdot z$, we obtain $a_1 = 1.59^{+0.11}_{-0.10} \times 10^{-10}~\mathrm{m/s^2}$, providing  evidence 
for a $z$-evolution. We find similar results using various dark matter halo profiles as well as the Modified Newtonian Dynamics framework in our 3D forward modelling.\\
\textit{Conclusions.} Our results show that the RAR persists at intermediate redshift, with statistically significant redshift evolution of the characteristic acceleration, pointing to a possible evolution of the baryon-missing mass connection over cosmic time.}

\maketitle



\setcounter{section}{0}
\section{Introduction}

The missing mass problem in galaxies has been studied since the 1930s (\citealt{bab}), but gained prominence in the second half of the 20th century (\citealt{vdh}), especially since the 1970s with observations of flat rotation curves (RCs) in disk galaxies (e.g., \citealt{Rubin}, \citealt{Bosma1}), deviating from the expected Keplerian decline. This mass discrepancy ($M_{\rm tot}/M_{\rm bar} \sim v_{\rm obs}^2 / v_{\rm bar}^2 \gg 1$, where $v_{\rm obs}$ and $v_{\rm bar}$ are the circular velocities, respectively observed vs. generated by the baryonic content of the galaxy) led to the hypothesis of dark matter (DM) halos surrounding galaxies (e.g., \citealt{White}, see also reviews such as \citealt{bull17}, \citealt{bosma}).

This problem was later also quantified in terms of the local radial acceleration in galaxies (\citealt{1990A&ARv...2....1S}, \citealt{McGaugh04}), which led to the identification of the apparently fundamental radial acceleration relation (RAR; \citealt{rar}, \citealt{lelli17}, \citealt{2023MNRAS.525.6130S}), that connects the observed acceleration $( a_{\rm{tot}}=v^2_{\rm obs}/r= |\partial \Phi_{tot}(r)/\partial r|$) to the gravitational acceleration generated by the baryons ($a_{\rm{bar}}=v^2_{\rm bar}/r = |\partial \Phi_{bar}(r)/\partial r|$) across all radii. The RAR has been empirically calibrated using the SPARC sample \citep{lelli16b}, and is usually parametrised by the following relation \citep{McGaugh08,Famaey1}:
\begin{equation}
a_{\rm tot} = \frac{a_{\rm bar}}{1 - \exp\big(-\sqrt{a_{\rm bar}/a_0}\big)} \label{RAR}.
\end{equation}
This relation exhibits a scatter of  $\sim0.11$~dex \citep{rar}, with a characteristic acceleration scale of $a_0 = 1.2 \pm 0.26 \times 10^{-10}~\rm m/s^2$, below which $a_{\rm bar}$ begins to deviate systematically from $a_{\rm tot}$. \cite{Desmond2} and \cite{2024MNRAS.530.1781D} refined this analysis on the SPARC sample, finding an even lower intrinsic scatter: $\quad \sigma_{\rm \emph{intrinsic}} = 0.034 \pm 0.001_{\rm stat} \pm 0.001_{\rm sys} \, \mathrm{dex}.$

A few  numerical studies have shown that a relation resembling the RAR  can occur in the \(\Lambda\)CDM framework using hydrodynamical simulations (\citealt{Keller}, 
 \citealt{Garaldi}, \citealt{Tenneti}, \citealt{Dutton}, \citealt{Mayer}) and semi-empirical analytic models  (\citealt{2000ApJ...534..146V}, \citealt{dc16}, \citealt{nav}, \citealt{Gru}, \citealt{Paranjape}, \citealt{li22}). Nevertheless, the detailed results of these studies  differ slightly from each other, so that it remains unclear whether some of them do actually reproduce in detail the properties of the observed RAR, including its functional form, very low scatter and the full diversity of RC shapes to which it applies \citep{Ghari, Desmond1}.  
 
Alternatively, the RAR may reflect a modification of gravity (or, more generally, of dynamics) rather than unseen matter, which would naturally lead to a very small scatter of the relation. Modified Newtonian Dynamics (MOND; \citealt{Milgrom1}, \citealt{Milgrom2}, \citealt{Milgrom3}; see also \citealt{Famaey} for a recent review) introduced a characteristic acceleration scale \( a_0 \sim 10^{-10} \:\:\rm{m/s^{-2}} \), below which Newtonian dynamics break down.  Remarkably, the MOND hypothesis predicted the existence and shape of the RAR decades before it was empirically established. 

In the standard MOND framework, \(a_0\) is typically considered a universal constant. However, since MOND is a non-relativistic theory, extending it to a cosmological context requires additional assumptions \citep[e.g.,][]{Bekenstein,Famaey1,Milgrom2020,Blanchet}. As a result, there is no clear consensus on whether \(a_0\) should remain constant or evolve with cosmic time in the MOND framework.  For example, some studies consider \(a_0\) as a constant parameter, while others propose that it might scale with the Hubble parameter, $H(z)$, or follow alternative evolutionary paths (e.g., \citealt{hof}, \citealt{Maeder}, \citealt{dchan}, \citealt{2024MNRAS.535L..13G}).

In contrast, within \(\Lambda\)CDM, it remains unclear whether the RAR can be fully reproduced, as it would have to arise from the complex interplay between baryonic and DM distributions, and any $z$-evolution of the relation would then reflect the evolving properties of these components.  In any case, the degree of $z$ evolution of $a_0$ in hydrodynamical simulations remains uncertain \citep[e.g.,][]{Garaldi,Tenneti,Mayer}, likely due to differences in the used feedback models.

While these theoretical frameworks offer different perspectives on the redshift evolution of the RAR, observational studies beyond \(z=0\) remain scarce--except for \citet{Varasteanu}, who investigated the RAR up to \(z \sim 0.08\), finding a larger $a_0$ value than at $z=0$. In this letter, we study the RAR at intermediate redshifts up to \(z \sim 1.44\), offering new insights into the evolution of \(a_0\) over cosmic time. 
This analysis is based on the results presented in \cite{paperI} (hereafter  \citetalias{paperI}), where we conducted a disk-halo decomposition using \textsc{GalPaK$^{\rm 3D}$} \citep{galpak} on a large sample of star-forming galaxies (SFGs), using deep data from the MUSE Hubble Ultra Deep Field survey (MHUDF, \citealt{udf2}) offering high signal-to-noise (S/N) data ranging from $\rm{S/N}\sim 10$ to $>$100.

\section{Sample selection and methodology}
\label{data}

To investigate the RAR  at intermediate $z$, we selected from the parent sample in \citetalias{paperI}, 79 SFGs with $M_\star>10^{8.8}$~\msun{}.  This sample has the same minimum stellar mass  (i.e. is mass-complete) over  $0.33<z< 1.44$ and is part of the MHUDF (\citealt{udf2}); further details are provided in Appendix~\ref{data_selection}. 

In \citetalias{paperI}, we performed a detailed 3D disk–halo decomposition on the parent sample of SFGs presented in Fig.~\ref{hist} (data points encircled in red), testing six different DM halo profiles. In the present study, we use the generalised profile of \citet[][hereafter DC14]{dc14} since they provided the best fits compared to other halo models, including the Navarro-Frenk-White profile (NFW, \citealt{nfw}). Note that \citet{dc16} showed that the RAR is more readily reproduced in models with mass-dependent DM density profiles (DC14) than with universal NFW profiles, which struggle to match the observed shape of the relation--particularly in low-$M_{\star}$ galaxies.
 
We compute $a_{\rm bar}$ and $a_{\rm tot}$ from the best-fitting DC14 disk–halo models, propagating uncertainties from the Markov Chain Monte Carlo (MCMC)-derived velocity components. To minimise resolution-driven biases, measurements within $r<2$~kpc (corresponding to $\sim$ one spatial resolution element, i.e. $\sim$0.2$^{\prime\prime}$) are excluded (see Appendix~\ref{methodology} for more details).

While our analysis relies on parametric modelling, we performed independent consistency checks of the disk--halo decomposition in \citetalias{paperI}, finding consistent results. Additionally, we assess potential systematic uncertainties in the stellar mass estimates by deriving stellar mass-to-light ratios for our sample. As discussed in Appendix~\ref{robsustness1}, these are lower than those assumed at $z=0$ in SPARC, but fully consistent with the expected decrease with $z$  from independent studies \citep{Drory}.

\section{Results}
\label{results}
\subsection{RAR at $0.33<z<1.44$}
\label{Res:MHUDF RAR}

\begin{figure}%

    \centering
    \includegraphics[width=7.4cm]{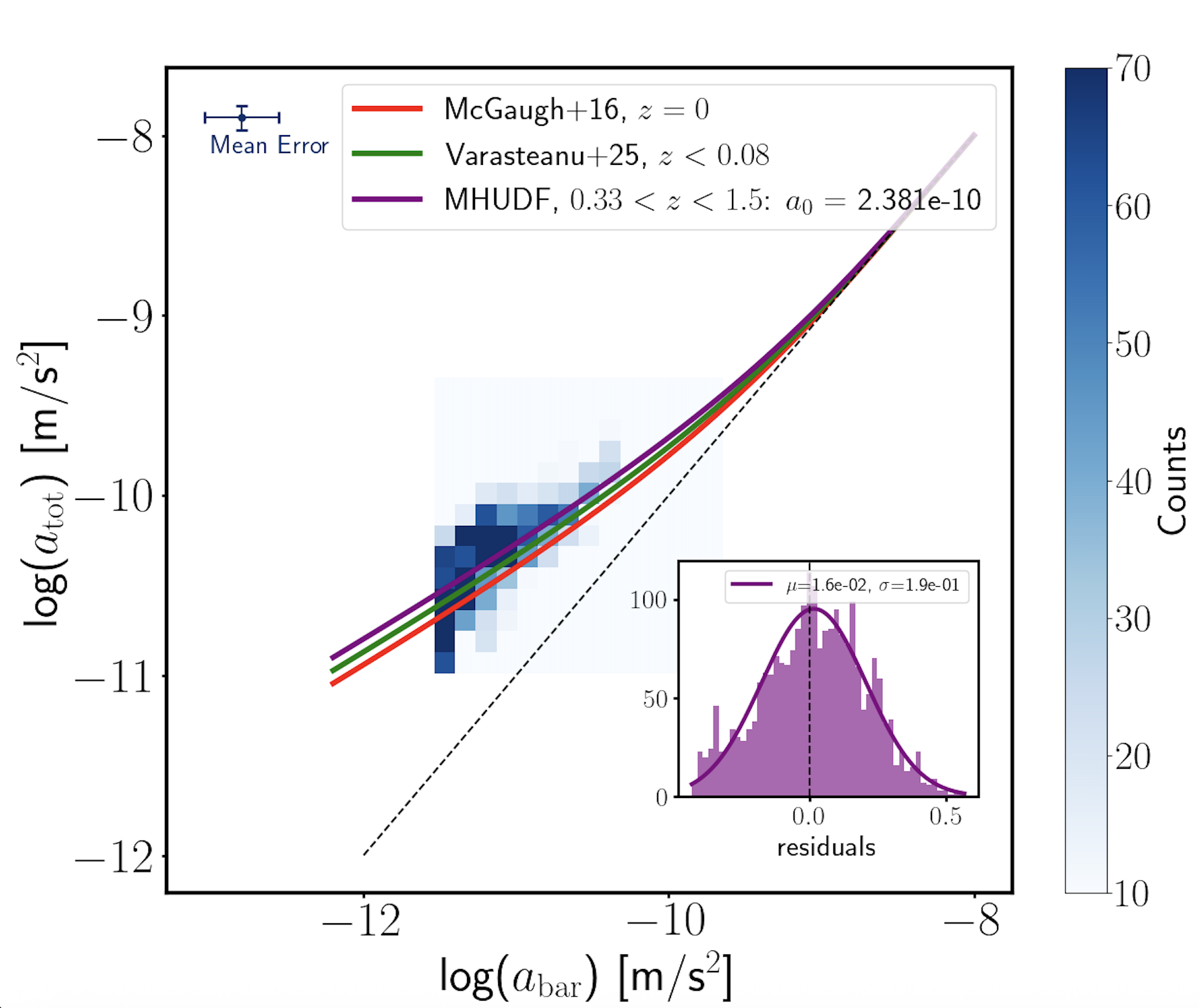}
  \caption{
RAR for the MHUDF sample.  
The purple curve shows the best-fit RAR  at $0.33 < z < 1.44$ using Eq. \ref{RAR}, while the red and green curves show the \cite{rar} relation at $z=0$ and the \cite{Varasteanu} relation at $z<0.08$, respectively. The black dotted line indicates the 1:1 relation.  The histogram inset displays the residuals with respect to our fit. The cross in the upper left corner indicates the mean $1\sigma$ uncertainties.}

    \label{rarr}%
\end{figure}

 Figure~\ref{rarr} presents the RAR at $0.33 < z < 1.44$,  using data points from 79 individual model RCs. As illustrated, the RAR at intermediate-$z$ is reminiscent of the RAR observed at $z=0$ \citep{rar}. Our sample, unlike SPARC, predominantly probes the low-acceleration regime, as it is largely composed of low-mass, DM-dominated SFGs lacking prominent bulges.
 
\begin{figure}
   \centering
    \includegraphics[width=0.44\textwidth,angle=0,clip=true]{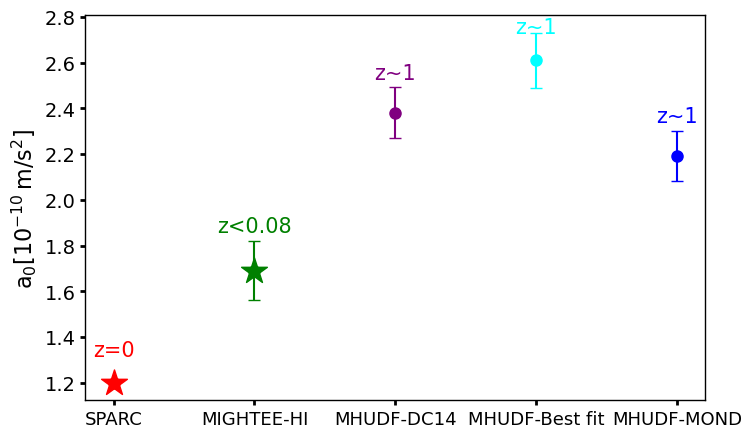}%
   \centering   
    \caption{Best-fit $a_0$ values obtained by fitting Eq.~\ref{RAR} on the full sample, using the resolved RAR tracks derived under different modelling assumptions. The purple data point shows $a_0|{z\sim1}$ obtained using the DC14 halo profile applied uniformly to the full sample (Sect.~\ref{Res:MHUDF RAR}). The cyan point shows $a_0|{z\sim1}$ derived from a galaxy-by-galaxy best-fitting $\Lambda$CDM model, where each galaxy is assigned the DM profile that maximises its Bayesian evidence (Appendix~\ref{RARothermodel}). The blue point shows $a_0|{z\sim1}$ recovered from the MOND-based framework (Appendix~\ref{RARMOND}). The red star indicates the fiducial $a_0|{z\sim0}$ from \cite{rar}, while the green star shows the result from \citet{Varasteanu}.}
\label{comp}
\end{figure}

We fit the data with equation \ref{RAR}, leaving $a_{0}$ as a free parameter. For the fits, we use the Python package \texttt{Roxy} \citep{roxy}, which implements the Marginalised Normal Regression (MNR) method. This approach simultaneously accounts for uncertainties in both $x$ and $y$, as well as for the intrinsic scatter in the relation. We apply the MNR to the accelerations in base 10 logarithmic space, and we adopt a uniform broad prior on $\log a_0$, ranging from $-15$ to $5$, and on  the intrinsic scatter between 0 and 3 dex. We show the best-fit curve in purple in Fig.~\ref{rarr},  and as the purple data point in Fig.~\ref{comp}, which corresponds to a characteristic acceleration scale of 
\begin{equation}
    a_0|_{z\sim1} = 2.38 ^{+0.12}_{-0.10} \times 10^{-10}~\mathrm{m/s^2}, \label{eq:main:a0}
\end{equation}
 with the errors denoting the $95\%$ confidence intervals (CI) from our MCMC fits. This value is significantly higher, by $\sim 19\sigma$, than the canonical value  of $a_0|_{z=0} = 1.2  \pm 0.26\times 10^{-10}~\mathrm{m/s^2}$ inferred for the SPARC sample (\citealt{rar}, shown as the red curve in Fig.~\ref{rarr}, and as the red star in Fig. ~\ref{comp}), and  by $\sim 5\sigma$ than the value of $a_0|_{z<0.08} = 1.69 \pm 0.13 \times 10^{-10}~\mathrm{m/s^2}$ inferred by \cite{Varasteanu} for the  MIGHTEE-HI sample (sown as the green curve in Fig.~\ref{rarr}, and as the green star in Fig. ~\ref{comp}). 
 At face-value, this would imply that disks are more submaximal at higher $z$ than at $z=0$ in agreement with the large DM fractions found in \citetalias{paperI}. 

The histogram in the inset of Fig.~\ref{rarr} shows the residuals with respect to the best-fit RAR at $0.33<z<1.44$, which are tightly peaked, with a mean offset of $\mu = 0.016$ dex and a standard deviation of $\sigma = 0.19$ dex, similar to the intrinsic scatter ($\sigma_{\rm intr.}\sim 0.17$ dex).  The scatter that we recover for the intermediate-$z$ sample is $\sim \times 1.5$ higher than the one inferred by \cite{rar} for the  SPARC sample, which is in the order of $\sim 0.11$ dex. The larger scatter that we measure is most likely related to the broad $z$ range that we probe, as discussed in the next section, as well as to the poorer data quality at higher $z$. 

As mentioned before, our analysis relies on the parametric models from \citetalias{paperI}, where we found that the DC14 density profile provides the best representation of the data. To assess the robustness of our results, we explore alternative DM profiles in Appendix~\ref{RARothermodel} and also refit the data within a self-consistent MOND framework in Appendix~\ref{RARMOND}. The resulting $a_0|_{z\sim1}$ values from these various assumptions are always larger than the $z=0$ value and agree within the errors with the results presented above, as shown in  Fig.~\ref{comp} by the cyan and blue data points.

\subsection{Evolution of the RAR with $z$}
\label{Res:evol}
To carefully trace the potential evolution of the RAR with cosmic time, we applied a quantile-based binning of the galaxy redshift distribution. Specifically, we divided the sample into four quantile intervals containing approximately the same number of data points and, for each interval, we fit the RAR with Eq. \ref{RAR} to extract the best-fit characteristic acceleration scale.  The fits are performed following the same procedure as in Sect.~\ref{Res:MHUDF RAR}. We adopted this equal-population binning to stabilise the Bayesian fits by ensuring comparable statistical weight in each bin; note that the bins correspond to adjacent, non-overlapping $z$ intervals. 

\begin{figure}[!htb]
\centering
        \includegraphics[width=8cm]{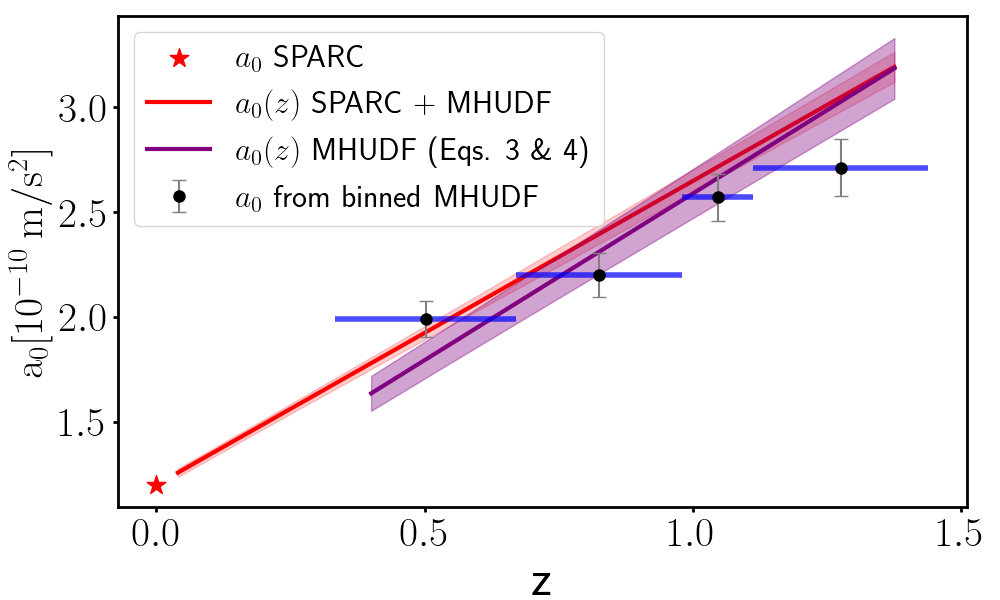}
   \caption{Redshift evolution of the characteristic acceleration scale. In each quantile-based $z$ bin (blue bars), we fit the RAR with Eq. \ref{RAR} to derive $a_0$, with uncertainties shown as grey error bars. The star marks the SPARC $z=0$ value. The purple line shows the predicted evolution, using the best-fitting parameters from the global, $z$-dependent fit to the RAR for the full sample (i.e using Eq. \ref{RAR} where $a_0$ was substituted with Eq. \ref{andreea}, - see Eq. \ref{eq:fits:a1}). The red line shows the same, but when including  SPARC  in the fit. The coloured shaded region shows the 1$\sigma$ errors for the fits. 
   }
    \label{rar_z}%
\end{figure}

We show the resulting $a_0(z)$ values in Fig.~\ref{rar_z} as the black data points.   As illustrated, the characteristic acceleration systematically increases with redshift, rising from $\sim 1.99\times10^{-10}~\mathrm{m/s^2}$ in the lowest $z$-bin to $2.71\times10^{-10}~\mathrm{m/s^2}$ in the highest.

We evaluated the intrinsic scatter in each redshift bin and found an increase with $z$, from $\sigma_{\rm intr.}=0.13$ dex in the lowest redshift bin--below the full-sample value--to $\sigma_{\rm intr.}=0.19$ dex in the highest. Nevertheless, the scatter in the individual  $z$-bins remains larger than that observed in the SPARC sample, likely reflecting the degradation in data quality and spatial resolution with increasing $z$, and the lower statistics.

To complement this binned analysis, we also performed a global, parametric fit to quantify the $z$-dependence of the RAR by introducing a $z$-dependent acceleration scale of the form: 
\begin{equation}
a_0(z) = a_{0}(0) + a_{1}\cdot z, \label{andreea}
\end{equation}
 where the coefficient $a_{1}$ captures the evolutionary trend with $z$, following the approach of \citet{Varasteanu}. We then refit the RAR for the entire sample using Eq.~\ref{RAR}, where we substitute $a_0$ with the $z$-dependent formulation (Eq. \ref{andreea}), treating both $a_{0}(0)$ and $a_{1}$ as free parameters in the fits.   We obtain: 
\begin{equation}
a_0(0) = 1.0^{+0.04}_{-0.04} \times 10^{-10}~\mathrm{m\,s^{-2}}, \quad
a_1 = 1.59^{+0.10}_{-0.10} \times 10^{-10}~\mathrm{m\,s^{-2}},
\label{eq:fits:a1}
\end{equation}
 with the errors denoting the $95\%$ CI. We note that including the SPARC sample \citep{lelli17} in the fits yields similar results.  

We show this linear relation, using the best-fit values of  $a_{0}(0)$ and $a_{1}$ from the fit to the MHUDF RAR (Fig.~\ref{rarr}) as the purple line in Fig.~\ref{rar_z}. The red line shows the same, but when the SPARC sample is included in the fit.  In  Appendix~\ref{appendix:corner}, we present the resulting posterior distributions of the fitted parameters, and discuss the small offsets between the binned estimates and the global fit.  

We emphasise that this linear parametrisation of $a_0(z)$  (Eq. \ref{andreea}) is intended as a simple, phenomenological description to quantify the presence and sign of a possible $z$-dependence rather than to provide a physically motivated description.  If the real $z$-dependence of $a_0$ is non-linear or more complex, the resulting evolutionary trend could be affected by systematic biases.  

In conclusion, the results presented in this section suggest that while the RAR persists out to $z \sim 1.44$, the value of $a_0$ increases with $z$, potentially reflecting changes in the baryon-DM coupling, in the feedback efficiency, or in modified gravity over cosmic time.

\section{Discussion}
\label{Discuss}

The results presented above indicate that the RAR evolves with $z$. 
We quantified this evolution with a simple, linear model  (Eq.~\ref{andreea}), obtaining  $a_1 = (1.59^{+0.11}_{-0.10}) \times 10^{-10}~\mathrm{m/s^2}$. Our result can be compared to the $z \leq 0.08$ study of \citet{Varasteanu}, who found 
\( a_1 = (4.47 \pm 1.88) \times 10^{-10}~\mathrm{m\,s^{-2}} \) from a smaller sample. While their analysis provided only mild evidence (\(\sim 2.4\sigma\)) for a $z$-evolution, our broader \(z\)-baseline and larger sample yield a more statistically significant detection (at a $\sim 30 \sigma$ level). The two measurements are statistically consistent within  \(\sim 1.5\sigma\). 

In the context of $\Lambda$CDM, our results show a roughly comparable $z$-trend as reported by \cite{Mayer} and \cite{Keller}, although the robustness of some of these results to reproduce the observed low-$z$ RAR has been questioned (\citealt{2016arXiv161007538M}). For example, \citet{Mayer} find that \(a_0\) grows by a factor of $\sim3$ from $z=0$ to $2$, slightly less than the factor of $\sim4$ increase inferred here over the same $z$ range.  
In the context of modified gravity, our measured $a_0(z)$ is faster than that of $H(z)$ \citep{Milgrom1}, and is in tension with the evolution predicted by the Scale Invariant Vacuum paradigm \citep{2024MNRAS.535L..13G}.
 
 An evolution of the RAR with $z$ should be reflected in the baryonic Tully–Fisher relation (bTFR). Current bTFR measurements based on \ion{H}{i} detections at $z \lesssim 0.4$ \citep{Ponomareva, Jarvis} show little to no evidence for evolution, but are limited by small samples, narrow $z$ baselines, and large uncertainties. Our $z$-dependent RAR fit (Eq.~\ref{eq:fits:a1}) predicts a bTFR zero-point shift of $\Delta{\rm ZP}\approx -0.2$ dex between $z\sim0$ and $z\sim1$.  At higher $z$, \citet{Ubler} find a decrease of the bTFR zero-point between $z\sim0$ and $z\sim0.9$ ($\Delta{\rm ZP}=-0.44$), though their $M_{\rm bar}$ estimates rely on scaling relations and neglect $M_{\rm \ion{H}{i}}$, while \cite{Alexandre} found no evolution of the bTFR at $z\sim1$, accounting for neutral gas from scaling relations. Looking ahead,  deep \ion{H}{i} surveys with the Square Kilometre Array will provide critical tests of the potential evolution of both the bTFR and the RAR with $z$.

\section{Conclusions}
\label{concl}

In this study, we analysed the RAR at intermediate redshifts ($0.33 < z < 1.44$), for a mass-complete sample of 79 SFGs with $8.8<\log(M_{\star}/M_{\odot})<11$,  using 3D forward modelling. To do so, we used the results obtained in \citetalias{paperI} from the disk-halo decomposition. Our analysis has yielded several key findings:
\begin{itemize}
 \item The RAR for our $z\sim1$ sample  exhibits a higher  acceleration scale $ a_0|_{z\sim1} = 2.38 ^{+0.12}_{-0.10} \times 10^{-10}~\mathrm{m/s^2}$  compared to the canonical value inferred by  \cite{rar} for the $z = 0$ SPARC sample (Fig.~\ref{rarr});
 \item  A larger value of $ a_0|_{z\sim1}$ is found regardless of the model used to derive the total and the baryonic accelerations (Fig.~\ref{comp}), including various $\Lambda$CDM and MOND models for self-consistency (Appendix \ref{RARothermodel} and \ref{RARMOND});
\item By fitting Eq.~\ref{RAR} independently in four redshift bins, we recover a systematic increase of the best-fit $a_0$ with $z$  (Fig.~\ref{rar_z});
\item Fitting Eq.~\ref{RAR} to the full sample by substituting $a_0$ with the redshift-dependent formulation, $a_0(z) = a_0(0) + a_1\cdot z$, yields $a_1 = (1.59^{+0.11}_{-0.10}) \times 10^{-10}~\mathrm{m/s^2}$, providing evidence for an increase of $a_0$ with redshift (Figs.~\ref{rar_z} and~\ref{corner}).
\end{itemize}
Some hydrodynamical simulations and modified-gravity frameworks predict a possible redshift evolution of $a_0$, although the magnitude and functional form vary across models. At present, theoretical expectations remain too diverse to make definitive comparisons. Our results, therefore, motivate future efforts to refine predictive models and explore the physical mechanisms that could shape any evolution of the RAR with cosmic time.

\begin{acknowledgements}
We thank the anonymous referee for constructive comments that improved the manuscript. JFr acknowledges stimulating discussions on the topic with S. Trujillo-Gomez.
This work is based on observations collected under ESO programmes 094.A-0289(B), 095.A-0010(A), 096.A-0045(A), 096.A-0045(B) and 1101.A-0127. This research made use of the following open source software: \texttt{Astropy} \citep{astropy}, \texttt{numpy} \citep{numpy} and \texttt{matplotlib} \citep{plot}. BC, NB and JFe acknowledge support from the  ANR DARK grant (ANR-22-CE31-0006). HD is supported by the Royal Society University Research Fellowship grant 211046.
\end{acknowledgements}


\bibliographystyle{aa}
\bibliography{references_fin}

\clearpage 

\begin{appendix}

\section{Sample of intermediate-$z$ SFGs}
\label{data_selection}

For this study, we use the sample of SFGs with RC decomposition from \citetalias{paperI}, which are drawn from the MHUDF survey \citep{udf2}.  A comprehensive description of the initial sample selection is provided in \citetalias{paperI}; here we briefly summarise the key criteria.  The parent sample consists of intermediate-$z$ SFGs with high-S/N ($\sim10$ to $>100$) in either $\rm{[O\textsc{ii}]}\:\lambda 3727$, $\rm{H\beta}$,  $\rm{[O\textsc{iii}]}\:\lambda 5007$,  $\rm{H\alpha}$, which are well resolved, have inclinations $i>30^\circ$, are rotationally supported ($v_{\rm max}/\sigma > 1$), and show no evidence for ongoing mergers.
For the present analysis, we further restrict the sample to galaxies classified as 'regular' in \citetalias{paperI}. This classification was based on a visual inspection of rest-frame optical/IR morphologies and kinematic maps, selecting systems showing no obvious signs of gravitational interactions (e.g. tidal features, asymmetries, or disturbed velocity fields). In addition, we only selected galaxies that exhibit small residuals in the kinematic modelling, i.e. which have good fits, quantified by $ \log(\mathcal{Z}) < 15000$, where  $\log(\mathcal{Z})$ is the model evidence.\footnote{We note that we rescale the evidence $\log(\mathcal{Z})$ by a factor of -2 so that it is on the same scale as the usual information criterion; in this convention, lower values correspond to better-fitting models.}

\begin{figure}
   \centering
    \includegraphics[width=0.45\textwidth,angle=0,clip=true]{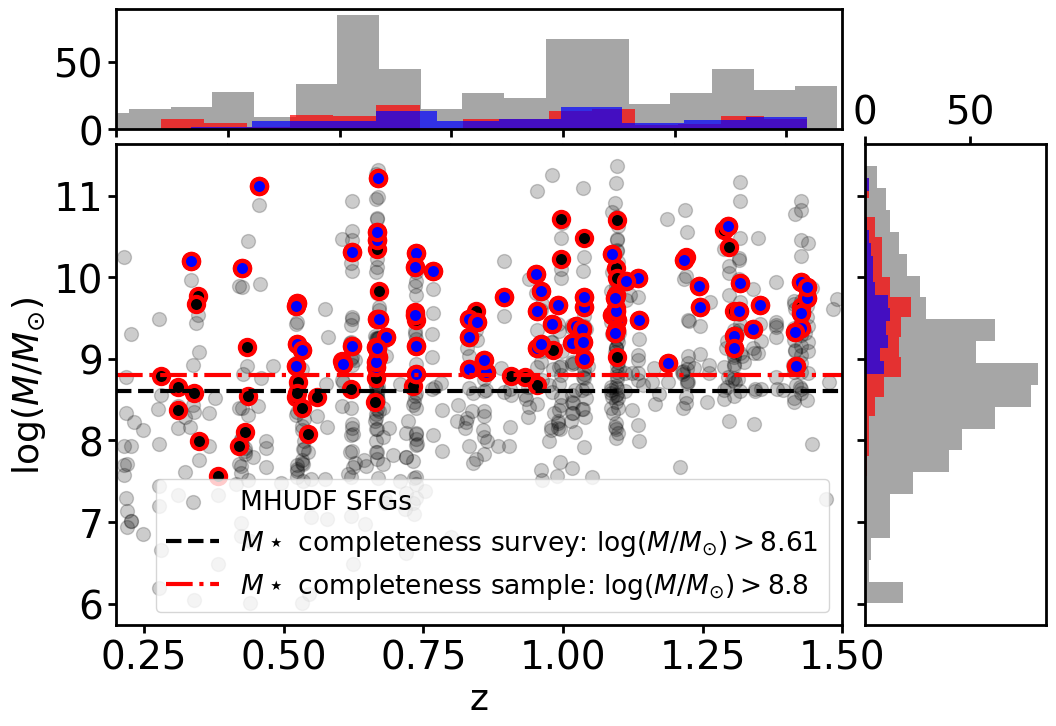}
   \centering   
  \caption{Stellar masses versus redshift for all MHUDF SFGs. 
The black dashed line indicates the mass completeness limit of the survey, which is reasonably complete for $0.2 < z < 1.5$ down to $\log(M_{\star}/M_{\odot}) = 8.6$. The red dashed line shows the completeness of the parent sample from \citetalias{paperI}. Grey points show all SFGs observed in the MHUDF survey, red-outlined points correspond to the galaxies analysed in \citetalias{paperI}, and blue points indicate the galaxies studied in this work. 
The histograms on the top and right display the redshift and stellar-mass distributions, respectively, colour-coded to match the main panel.}

\label{hist}
\end{figure}

To robustly investigate the redshift evolution of the RAR, it is crucial to work with a sample defined by a uniform stellar-mass distribution across the redshift range considered, i.e. imposing the same minimum $M_\star$ at all redshifts.  Figure~\ref{hist} indicates that the MHUDF survey is complete down to a stellar-mass of $\log(M_{\star}/M_{\odot})=8.6$ over  $0.2<z<1.44$, whereas the sample from \citetalias{paperI} is complete to $\log(M_{\star}/M_{\odot})=8.8$ due to the additional S/N criteria that were imposed.  We therefore adopt a stellar-mass cut of $\log(M_{\star}/M_{\odot}) > 8.8$, yielding a final sample of 79 SFGs (blue-red data points in Fig.~\ref{hist}) used in this study. These galaxies span stellar masses and redshifts in the ranges $8.8 < \log(M_{\star}/M_{\odot}) < 11$ and $0.33 < z < 1.44$, respectively.

\section{Methodology}
\label{methodology}
 Based on disk-halo decomposition from  \citetalias{paperI}:
\begin{equation}
v_{\rm{c}}^2(r) = v_{\rm{DM}}^2(r) + v_{\rm{disk}}^2(r) + v_{\rm{HI}}(r) |v_{\rm{HI}}(r)| ( + v_{\rm{bulge}}^2(r)),
\end{equation} 
  we calculate the baryonic acceleration due to the baryonic mass distribution at every radius as: 
\begin{equation}
a_{bar} (r) = v_{\rm{disk}}(r)^2 /r + v_{\rm{HI}}(r)^2 /r \:(+  v_{\rm{bulge}}(r)^2 /r)
\label{v}
\end{equation}
and the total acceleration from:
\begin{equation}
a_{tot}(r) = v_{\rm{c}}(r)^2 /r.
\label{v}
\end{equation}

Here, $v_{\rm disk}(r)$, $v_{\rm HI}(r)$, and $v_{\rm bulge}(r)$ denote the contributions of the stellar disk, neutral atomic gas, and bulge\footnote{For the 4 galaxies in this sample which have a bulge-to-total ratio $B/T>0.2$, a bulge component is included in the disk--halo decomposition.} to the RC, respectively, while  $v_{\rm DM}(r)$ represents the contribution of the DM halo.  The circular velocity, $v_{\rm c}(r)$, is corrected for pressure support (asymmetric drift) following \citet{Dalcanton}, such that
$v_{\rm c}^2(r) = v_{\perp}^2(r) + v_{\rm AD}^2(r)$,
where $v_{\perp}(r)$ is the observed rotation velocity and $v_{\rm AD}(r)$ is the pressure support correction. We note that all velocity terms are intrinsic, i.e. corrected for any instrumental effects, including seeing.

We summarise here in brief the main ingredients of the 3D disk--halo decomposition (see \citetalias{paperI} for full details).  The DM halo is parametrised using the best-fitting DM density profile from \citetalias{paperI}, namely DC14 \citep{dc14}.  For the neutral gas component, we use a parametric model which assumes a constant \ion{H}{i} surface mass density, leading to $v(r)\propto\sqrt{\Sigma_{\rm HI}r}$.  The bulge, if present, is modelled as a \citet{Hernquist} spheroidal component.  The disk contribution is parametrised from the observed ionised gas surface brightness profiles using a Multi-Gaussian Expansion modelling  \citep{MGE}, whereas the normalisation of $v_{\rm disk}(r)$ is given by $M_{\star}$. Importantly, $M_{\star}$ itself is not fixed by photometric priors. Instead, it is dynamically inferred in our 3D forward modelling. Within the \cite{dc14} framework, the disk-halo degeneracy is mitigated by fitting for $\log(M_{\star}/M_{\rm halo})$ together with the virial velocity. This allows the stellar mass to be constrained directly by the kinematics. 

\begin{figure*}
   \centering
    \includegraphics[width=0.8\textwidth,angle=0,clip=true]{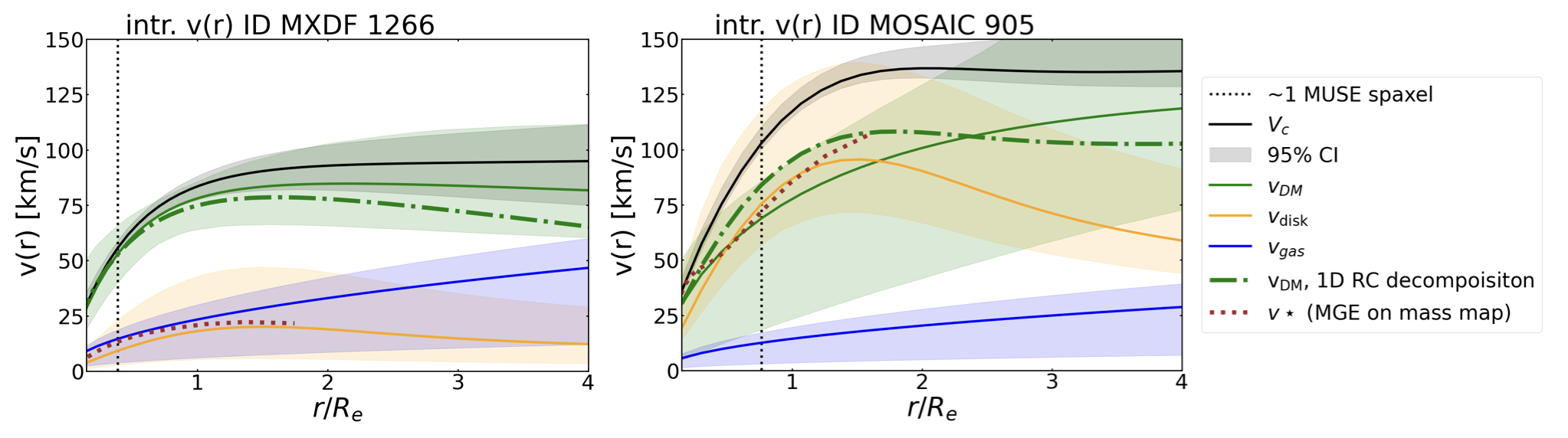}
 \centering
\caption{Examples of the 3D disk-halo decomposition for the two galaxies from the sample. 
The solid black line represents the circular velocity $v_{\rm{c}}(r)$ corrected for pressure support. The green curve represents the RC of the DM component. The solid orange and blue curves represent the disk and cold gas components, respectively.  The light shaded regions show the 95\% confidence intervals.  The dotted   brown line represents the stellar component obtained using the Multi-Gaussian Expansion modelling of resolved stellar mass maps obtained from pixel-by-pixel SED fitting, while the dashed-dotted green curve shows the DM component obtained independently from a 1D disk-halo decomposition.  All velocities are intrinsic, i.e. corrected for inclination and instrumental effects, including seeing (PSF). 
  The black dotted line shows the physical extent of a MUSE spaxel. }
\label{RCdecomp}
\end{figure*}

For illustration, we show in Fig.~\ref{RCdecomp}  the 3D disk--halo decomposition for two galaxies from our sample. We note that all catalogues and data products from our disk–halo decomposition, including the RCs, can be found on the DARK website\footnote{ \url{https://dark.univ-lyon1.fr/data-releases/}.}.
For MXDF~1266 (left, $M_{\star, \rm{SED}} = 10^{9.05}\, M_\odot$), the RC is dominated by the DM component, indicating a strongly submaximal disk compared to local higher-mass spirals (e.g., \citealt{lelli16b}).  On average, we find that most of our sample has DM  fractions larger than 50\% within $R_e$ (i.e. they are submaximal), as shown in \citetalias{paperI}. In contrast, the higher-mass galaxy, MOSAIC~905 (right,  $M_{\star, \rm{SED}} = 10^{10.3}\,M_\odot$), is dominated by the stellar component within $2R_e$.  

The $1\sigma$ errors in the velocity components are derived from the 68\% confidence intervals of the MCMC chains. Specifically, for each model parameter, we compute the 16th and 84th percentiles of the posterior distributions, providing the lower and upper bounds of the confidence interval. To estimate the uncertainties in the baryonic and total accelerations, we propagate the errors from the velocity components using standard error propagation techniques. 

To minimise the impact of observational uncertainties, particularly in regions where the rotation and acceleration profiles are poorly resolved, we exclude measurements in the innermost regions ($r < 2$ kpc) from the final fits to the RAR. Due to the finite spatial resolution of the MUSE instrument (0.2\arcsec spaxels, black dotted lines in Fig. \ref{RCdecomp}), measurements in these innermost regions are affected by resolution limitations that might bias our final results. A similar approach was taken in \cite{Varasteanu}. We note, however, that excluding measurements within the central 2 kpc for both the MHUDF and SPARC \citep{lelli17} samples changes the inferred $a_0$ by less than 10\%.

\section{Robustness tests and systematic uncertainties on disk masses}
\label{robsustness1}
In this appendix, we assess the robustness of our results with respect to the main modelling assumptions. We emphasise that our measurement of the RAR at intermediate $z$ is not strictly identical to the traditional local-Universe approach, as it relies on 3D forward modelling. While this technique represents the state of the art for intermediate-to-high-$z$ galaxy dynamics, it necessarily introduces a degree of model dependence, stemming from assumptions about the halo profile, disk geometry, gas distribution, etc. In this framework, both the total acceleration, $a_{\rm tot}(r)$, and the baryonic acceleration, $a_{\rm bar}(r)$, are derived from the same underlying model rather than independently, as is typically done at $z\sim0$.

Nonetheless, we note that in \citetalias{paperI}, we presented a series of consistency checks supporting the robustness of our disk--halo decomposition. Specifically:
(i) The total circular velocity, $v_{\rm{c}}(r)$, was independently measured using the traditional 2D line-fitting algorithm \texttt{CAMEL} \citep{camel}, yielding results in good agreement with those obtained from the 3D parametric modelling (rendering $v_{\rm c}(r)$ effectively model-independent).
(ii) The stellar disk contribution, $v_{\rm disk}(r)$, was independently derived via a Multi-Gaussian Expansion \citep{MGE} modelling applied to resolved stellar-mass maps obtained from pixel-by-pixel SED fitting across \(\sim19\) Hubble Space Telescope and James Webb Space Telescope bands, again showing good agreement (as illustrated by the brown dotted curves in Fig. \ref{RCdecomp}). 
(iii) For the unknown neutral gas contribution, $v_{\rm HI}(r)$, we explored a range of plausible \ion{H}{i} surface-density profiles--given their lack of direct constraints--and found that our results are consistent across various assumed profile shapes. 
(iv) Finally, we performed a fully independent 1D disk--halo decomposition by computing the baryonic contribution to the RC. This was achieved by solving the Poisson equation for the surface-density distribution of the stellar and cold gas components (i.e. by solving Eq.~4 of \citealt{Casertano}). This 1D approach yielded results consistent with those from the  3D modelling (as illustrated by the green dash-dotted curves in Fig. \ref{RCdecomp}).

\begin{figure*}
    \centering
    \includegraphics[width=0.55\textwidth,angle=0,clip=true]{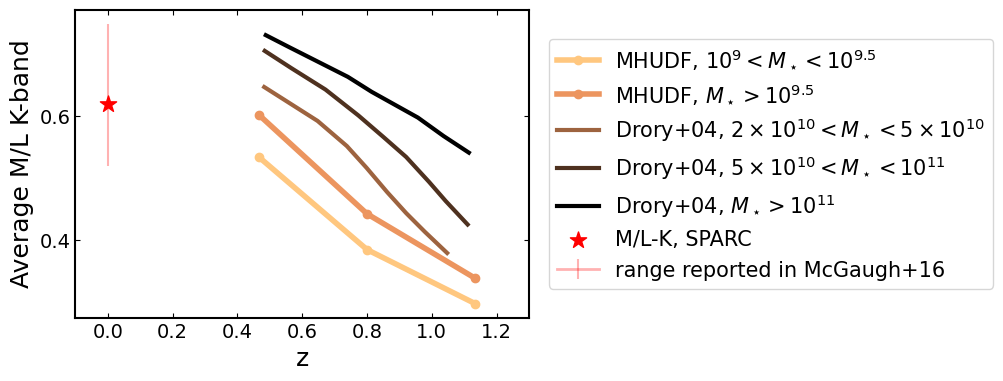}
 \centering
   \centering
    \includegraphics[width=0.35\textwidth,angle=0,clip=true]{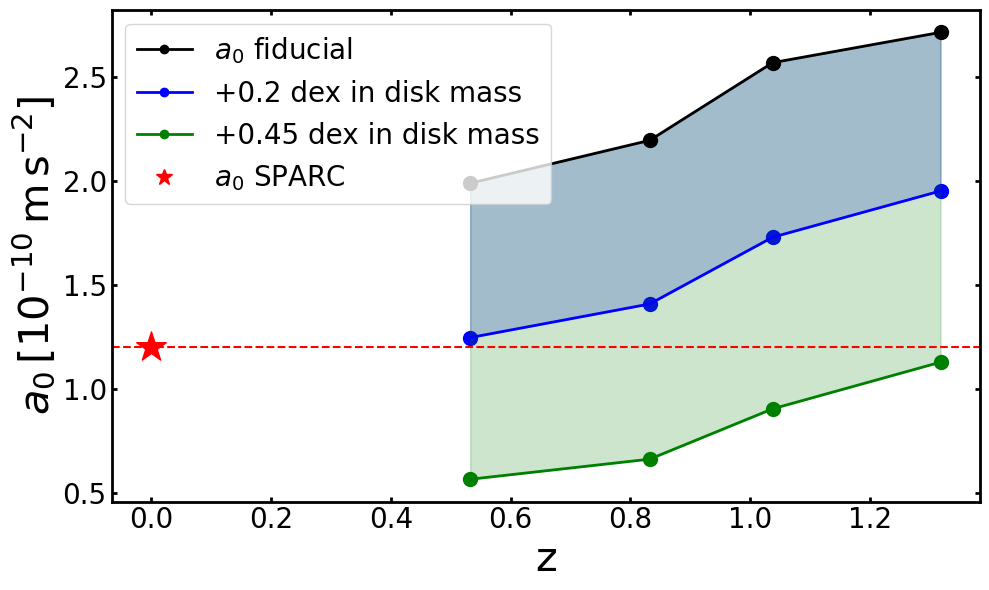}
 \centering
\caption{Left: Average mass-to-light ratios (M/L) in the K-band as a function of redshift for the MHUDF sample split into two-$M_\star$ bins (low-$M_\star$ bin-yellow, high-$M_\star$ bin-orange), along with the evolution inferred for a  sample of more than 5000 K-selected galaxies in \cite{Drory}. The red star shows the M/L-K ratio adopted for the SPARC sample. Right: Impact of systematic disk mass offsets on the inferred $a_0(z)$. The black points correspond to the fiducial fits shown in Fig \ref{rar_z}. The blue points show the inferred $a_0(z)$  assuming a global shift of $+0.2$ dex in the mass budget of the sample, while the green points show the same, but when assuming a global shift of $+0.45$ dex . The red star and dotted lines show the canonical $a_0(z=0)$ inferred for the SPARC sample.}
 \label{mstarbias}
\end{figure*}

Given the central role of $M_{\star}$ in our results, it is important to note the dynamically
inferred stellar masses are in good agreement with independent SED-based estimates (Fig.~11 of \citetalias{paperI}), showing that the potential systematic uncertainties in $M_\star$ are negligible.
In addition, we compute the rest-frame K-band mass-to-light ratios ($M/L_K$) of our sample, using photometric data from the  Hubble Space Telescope and James Webb Space Telescope  (see \citetalias{paperI}), and compare them to independent measurements at similar redshifts
from \citet{Drory}.
Fig.~\ref{mstarbias}(left) shows that the $M/L_K$ of our sample are fully consistent with typical values reported at higher redshifts accounting for the mass dependence, and  are systematically lower than those adopted for SPARC at $z=0$, which have $M/L_K\sim 0.6$ in the K-band (corresponding to $M/L\sim 0.5$ at 3.6~$\mu$m,   \citealt{rar}).

To further assess the impact of systematic uncertainties in the disk mass estimates on the RAR, we also explore how global systematic shifts in the $M_\star$   would affect the inferred $a_0(z)$. Specifically, we recompute the RAR after applying uniform offsets to $M_\star$ across the sample. We find that reconciling our measurements with the canonical value $a_0(0) = 1.2 \times 10^{-10}\,\mathrm{m\,s^{-2}}$ at all redshifts would require systematically larger stellar masses, with offsets ranging from $\sim +0.2$ dex in the lowest-redshift bins to $\sim +0.45$ dex in the highest (Fig.~\ref{mstarbias}, right). As argued above and discussed in  \citetalias{paperI}, such large shifts are not supported by independent consistency checks.

We note that in order to improve our modelling, direct constraints on  the  gas components will be required. However, as discussed in  \citetalias{paperI}, given typical molecular gas fractions of $\sim 30$-50\% at $z \sim 1$ \citep{fre2}, this would introduce a systematic uncertainty of $\sim 0.2$ dex in the total disk mass.

Finally, in appendices \ref{RARothermodel} and \ref{RARMOND}, we explored the RAR from RC decompositions using different parametric models in the $\Lambda$CDM and MOND frameworks, respectively, finding consistent results with those presented in the main text. Together, these tests demonstrate that our results are not driven by any single modelling assumption, but instead, most likely, reflect a genuine physical trend.

\section{Robustness tests - RAR from different DM halo models in $\Lambda$CDM}
\label{RARothermodel}

To ensure that our conclusions on the redshift evolution of the RAR are not driven by methodological choices, we investigated how variations in our modelling assumptions influence the recovered accelerations $a_{\rm bar}(r)$ and $a_{\rm tot}(r)$. In \citetalias{paperI} we tested six DM density profiles in our disk–halo decomposition: (1) DC14 \citep{dc14}, which links the halo profile shape to the stellar-to-halo mass ratio; (2) NFW \citep{nfw}; (3) Burkert \citep{Burkert}; (4) Dekel–Zhao \citep{fre}; (5) Einasto \citep{einasto}; and (6) coreNFW \citep{read}. Our analysis showed that DC14 provides the best population-level fit, and we refer the reader to the Bayesian model comparison in \citetalias{paperI} for details. 

\begin{figure}
   \centering
    \includegraphics[width=0.4\textwidth,angle=0,clip=true]{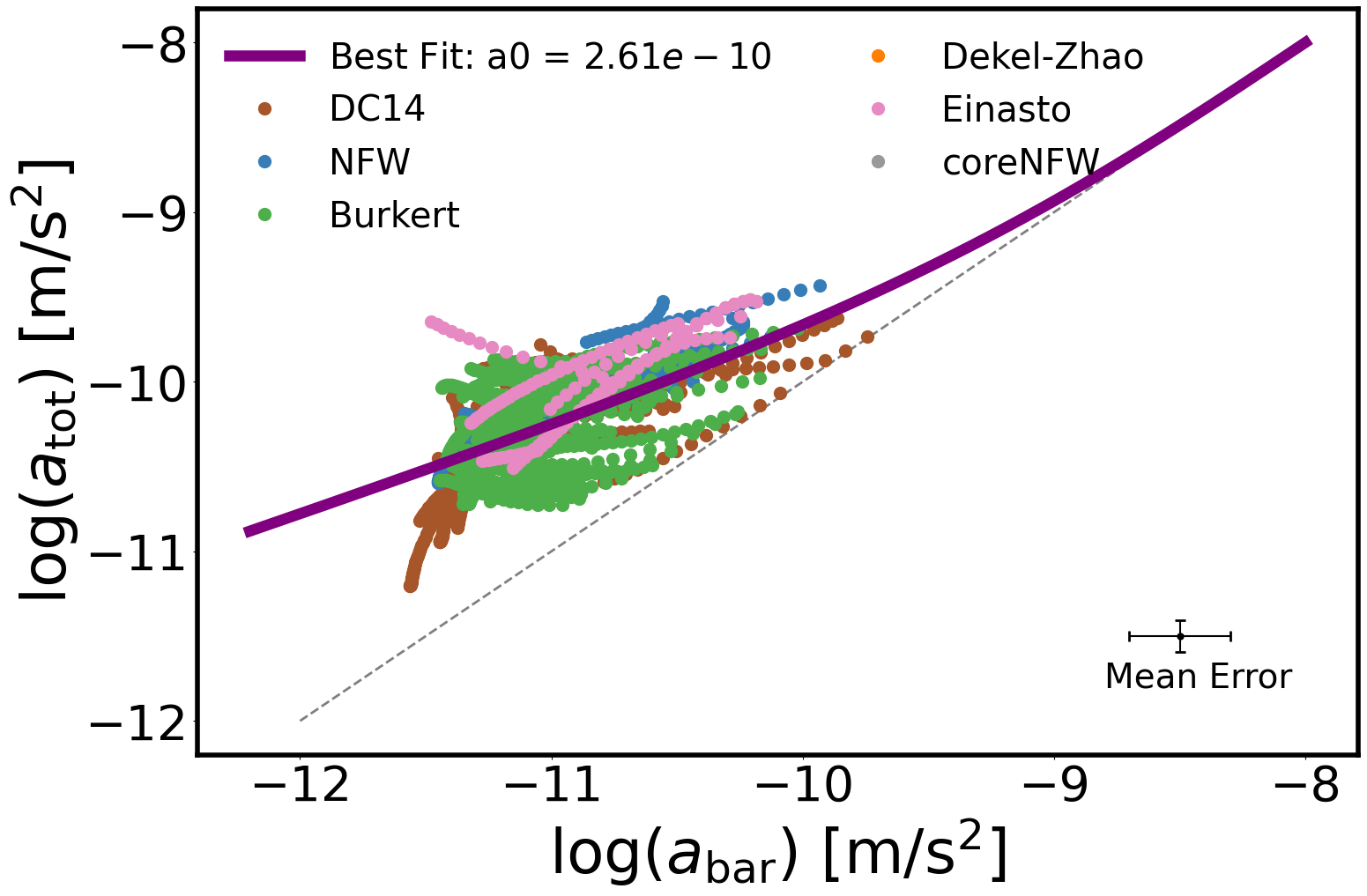}
 \centering
\caption{RAR of our intermediate-$z$ sample, with $a_{\rm bar}$ and $a_{\rm tot}$ derived from the model RCs corresponding to each galaxy's best-fitting $\Lambda$CDM halo profile. Colored points indicate the halo profile used in the disk–halo decomposition. 
The cross in the lower right corner indicates the mean $1\sigma$ uncertainties on the data points.
The dotted black line shows the 1:1 relation. The purple curve represents the best-fit RAR, with $a_0|_{z\sim1} = 2.61^{+0.13}_{-0.09} \times 10^{-10}~\mathrm{m/s^2}$ (i.e. as in Fig. \ref{rarr}).}

\label{RAR_bestfit}
\end{figure}

In this section, we adopt a galaxy-by-galaxy approach: for each system, we select the profile that yields the best Bayesian evidence in the pairwise comparisons. Each galaxy is therefore assigned its own preferred model--DC14 for some, NFW for others, etc. Using these best-fitting models for each galaxy, we recompute $a_{\rm bar}(r)$ and $a_{\rm tot}(r)$ (as detailed in Appendix~\ref{methodology}).  The resulting RAR tracks are shown as colored points in Fig.~\ref{RAR_bestfit}. We then refit the RAR for the full sample (Eq.~\ref{RAR}), treating the characteristic acceleration scale as a free parameter (as in Sect. \ref{Res:MHUDF RAR}); the best-fit relation is shown as the purple curve.  Additionally, in Fig.~\ref{comp}, we also show the best-fit value of $a_0|_{z\sim1}$ obtained in this framework as the cyan data point. We find:
\begin{eqnarray}
 a_0|_{z\sim1}= 2.61^{+0.13}_{-0.09} \times 10^{-10}~\mathrm{m/s^2},
 \end{eqnarray}
 a result that is consistent, within the errors, with the value derived in the main text (Eq.~\ref{eq:main:a0}), where the DC14 profile is applied uniformly to the full sample  (purple data point in Fig.~\ref{comp}). Fig.~\ref{comp} illustrates that regardless of the model used to derive the total and the baryonic accelerations, the  $ a_0|_{z\sim1}$ value inferred from the RAR is offset from the $z=0$ canonical value.

Following the same procedure presented in Sect.~\ref{Res:evol}, we fit the RAR in this framework using the $z$-dependent acceleration scale (Eq. \ref{RAR} substituting $a_0$ with Eq. \ref{andreea}), and obtain 
\begin{equation}
a_0(0) = 1.05 ^{+0.05}_{-0.05} \times 10^{-10} ~\mathrm{m/s^2}$;\: $a_1 = 1.63^{+0.13}_{-0.12} \times 10^{-10}~\mathrm{m/s^2},
\end{equation} which agrees within $1\sigma$ with the results presented in the main text when using DC14 uniformly for the whole sample.

Overall, these tests demonstrate that our results are fairly robust against the choice of DM profile: whether using DC14 uniformly or adopting the best-fitting model for each galaxy individually, the recovered redshift evolution of the RAR remains consistent. In the next section, we discuss the RAR derived from the MOND framework.

\section{Robustness tests - RAR from MOND models}
\label{RARMOND}

In order to investigate the RAR in the MOND framework, we first need to refit our MHUDF data directly with MOND parametric models, and then re-analyse the RAR from these new fits as in Sect.~\ref{results}. To refit our data, we extended our 3D modelling framework,  \textsc{GalPaK$^{\rm 3D}$} \citep{galpak},  to  MOND  (\citealt{Milgrom1})  \footnote{The MOND prescription used here is in isolation, without accounting for the external field effect (EFE), which arises when a system is subject to a constant external gravitational field.  Since our sample excludes interacting galaxies and cluster members, the EFE is expected to have only a minor impact.} using the interpolation functions defined in Sect.~3.3 of \cite{Famaey}. Specifically, we used two interpolation function families (\citealt{Milgrom3}, \citealt{rar}):
\begin{equation}
\rm{n-family}: \nu_n(x) = \left[ \frac{1 + \sqrt{1 + 4 (x / a_0)^{-n}}}{2} \right]^{1/n}, \label{eq:E1}
\end{equation}
\begin{equation}
\rm{\delta-family:} \quad \nu_\delta(x) = \left( 1 - exp \left(-(x/a_0)^{\delta/2} \right) \right)^{-1/\delta} \label{eq:E2} 
\end{equation}
which relate the total and baryonic accelerations as:
\begin{equation}a_{\rm tot} = \nu(a_{\rm bar}) a_{\rm bar}.\label{eq:atot}
\end{equation} 

Briefly, in our implementation within \textsc{GalPaK$^{\rm 3D}$}, the Newtonian baryonic acceleration $a_{\rm bar}(r)$ is computed directly from the disk, gas and, where applicable, bulge parametric models described in \citetalias{paperI} (and briefly detailed in Appendix \ref{methodology}, and thus, does not depend on any MOND-specific parameters). The code then evaluates the total acceleration $a_{\rm tot}(r)$ using Eq.~\ref{eq:atot}. 
Again, corrections for pressure support, following \cite{Dalcanton}, were applied to the RCs. 

The free parameters of the model are the same as in \citetalias{paperI}, with the addition of the MOND-specific parameters, namely $n$ or $\delta$, depending on the interpolation function, and the acceleration scale, $a_0$. We adopt the same priors as in \citetalias{paperI} for all shared parameters. For the \(n\)-family model (Eq.~\ref{eq:E1}), we use a flat prior $0.7 < n < 1.3$ , while for the $\delta$-family model (Eq.~\ref{eq:E2}), we use a similar prior: $0.7 < \delta< 1.3$. These values are motivated by   \cite{2024MNRAS.530.1781D}, who showed that the best fits to galaxy RCs were systematically close to $n \sim1$ or $\delta \sim 1$. Note that $\delta=1$ immediately gives back the usual form of Eq.~\ref{RAR}. \footnote{ We note that we fitted our data with MOND models in which the $\delta$ parameter was fixed to  $\delta=1$ or $\delta=2.5$, or allowed to vary with broad Gaussian priors, all of which yielded very similar results for our sample.}
We  find that the two MOND models (\ref{eq:E1} and \ref{eq:E2}) perform similarly well, with a Bayes factor $-2<\rm{\Delta log(\mathcal{Z})}<2$, meaning that they are indistinguishable,
and we therefore quote results for the $\delta$-family (Eq.~\ref{eq:E2}, with $\delta\sim1$, i.e. effectively Eq. \ref{RAR}) in the remainder of this Appendix.

Before discussing the RAR in the MOND framework, we attempt to quantify the redshift evolution of $a_0(z)$ from the individual fits to our MUSE data cubes
with: (i) $a_0$ free in each galaxy, with a broad flat prior $10^{-11}<a_0<10^{-9}$; (ii) $a_0$ fixed to the $z=0$ canonical value; and (iii) with the redshift evolution $a_0(z)$ (Eq.~\ref{andreea}),
where $a_1$ is a free parameter for each galaxy (with a broad flat prior $-1 \times 10^{-9}<a_1<1 \times 10^{-9}$) and $a_0(0)$ is kept fixed to the $z=0$ value.

Regarding (i),  $a_0$ was treated as a free parameter in the kinematic modelling of the individual MUSE cubes, and
we do not recover a single, well-defined characteristic acceleration for the sample. Instead, the inferred $a_0|_{z\sim1}$ values display broad galaxy-to-galaxy variations, ranging from $1.3\times10^{-11}$ to $9.6\times10^{-10}~\mathrm{m/s^{2}}$, with a median value of $a_0|_{z\sim1} = 2.27 \times 10^{-10}~\mathrm{m/s^2}$ (close to the value inferred in the main text in the $\Lambda$CDM framework - Eq.~\ref{eq:main:a0}). 
It is worth noting that similar results were obtained by \cite{noMOND}, who used Bayesian inference on the SPARC and THINGS samples to show that the probability of a single, universal acceleration scale common to all galaxies is effectively zero (but see also \cite{2018NatAs...2..924M} on the role of inclination and M/L uncertainties and priors on this statistical result).

We tested whether the acceleration scales inferred from our fits to the individual data cubes are consistent with being at or below the canonical SPARC value, 
by performing a one-sided  Kolmogorov–Smirnov (KS)  test. The fiducial value was modelled as a Gaussian distribution centred on  $a_0 = 1.2 \times10^{-10}~\mathrm{m/s^2}$ with a width equal to the measurement error $0.26 \times10^{-10}$ reported in \cite{rar}.  Our measurement uncertainties were incorporated by generating 10000 Monte Carlo realisations of the sample. In every realisation, the null hypothesis is rejected at a confidence level of 99\%, with a median statistic of $D =0.55$ and a median $p-\rm{value}=5.5\cdot10^{-20}$, indicating that the inferred values are systematically higher than the canonical $a_0(0)$ value. 

 Regarding (ii), when $a_0$ is fixed to $a_0(0)= 1.2\times 10^{-10}~\rm m/s^2$ in the kinematic modelling,  we found that the fits are worse, with the Bayes factor indicating positive evidence ($\rm{\Delta log(\mathcal{Z})}<-2$) for the models with $a_0$ as a free parameter. 
 We find that the model with free $a_0$ (i.e. model (i)) performs as well or better than that for a fixed $a_0$ (i.e. model (ii)) for $\gtrsim$ 70\% of the sample (similar results were obtained when comparing the model with $z$-varying $a_0$  (i.e. model (iii))  to the one with fixed $a_0$). We also note that the dynamically inferred $M_\star$ using the model with fixed $a_0$ are overestimated by $\sim 0.28$~dex with respect to the ones obtained in the main text using DC14, and are likewise elevated compared to independent SED-based estimates (although it does bring the M/L closer to those assumed in the SPARC database at $z=0$).

\begin{figure}
   \centering
    \includegraphics[width=0.35\textwidth,angle=0,clip=true]{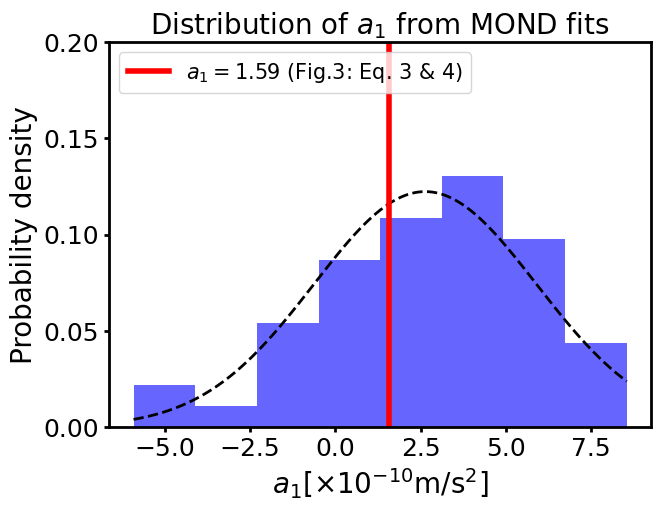}%
   \centering   
  
    \caption{Distribution of the MOND acceleration parameter $a_1$.    The purple histogram shows the recovered $a_1$ values from individual galaxy fits. 
    The vertical red line marks the reference value obtained from the RAR in the main text, $a_1 = (1.59^{+0.11}_{-0.10}) \times 10^{-10}~\mathrm{m/s^2}$, using the DC14 framework (Eq. \ref{eq:fits:a1}).} 
     
\label{monda1}
\end{figure}

Regarding (iii), we show the resulting  $a_1$ values from the individual fits in Figure~\ref{monda1} (solid histogram). 
This Figure reveals that the inferred $a_1$  values are predominantly positive, suggesting a positive correlation between the characteristic acceleration scale and redshift, 
and that the peak of the $a_1$ distribution is near the value (vertical line) obtained with the $z$-dependent fit to the RAR discussed in the main text (see Sect. \ref{Res:evol}: Eq.~\ref{eq:fits:a1}, Fig.~\ref{rar_z}). While individual $a_1$ values carry measurement uncertainties, we quantify the significance of the overall positive trend using a one-sample KS test (described below), which incorporates these uncertainties.

 
To quantify the statistical significance of this apparent offset from zero, we tested whether the $a_1$ values are consistent with being drawn from a zero-mean Gaussian distribution. This was achieved by performing a one-sample KS test against a standard normal null cumulative distribution function.  Our measurement uncertainties were incorporated by generating 10000 Monte Carlo realisations of the sample. In every realisation, the null hypothesis is rejected at the 99\% confidence level, with a median KS statistic of $D = 0.30$ and a median $p$-value of $1.2\times10^{-5}$. This indicates that the distribution of $a_1$ is inconsistent with a zero-mean Gaussian.

Taken together, these results indicate that the individual fits of our intermediate-$z$ galaxies do favour a positive $a_1$ value when fitted in a MOND framework, which agrees with the evolution inferred from the RAR when the data is fitted with DM halos (Sect.~\ref{Res:evol}), hinting at a positive correlation between the characteristic acceleration scale and redshift.

\begin{figure}
   \centering
    \includegraphics[width=0.4\textwidth,angle=0,clip=true]{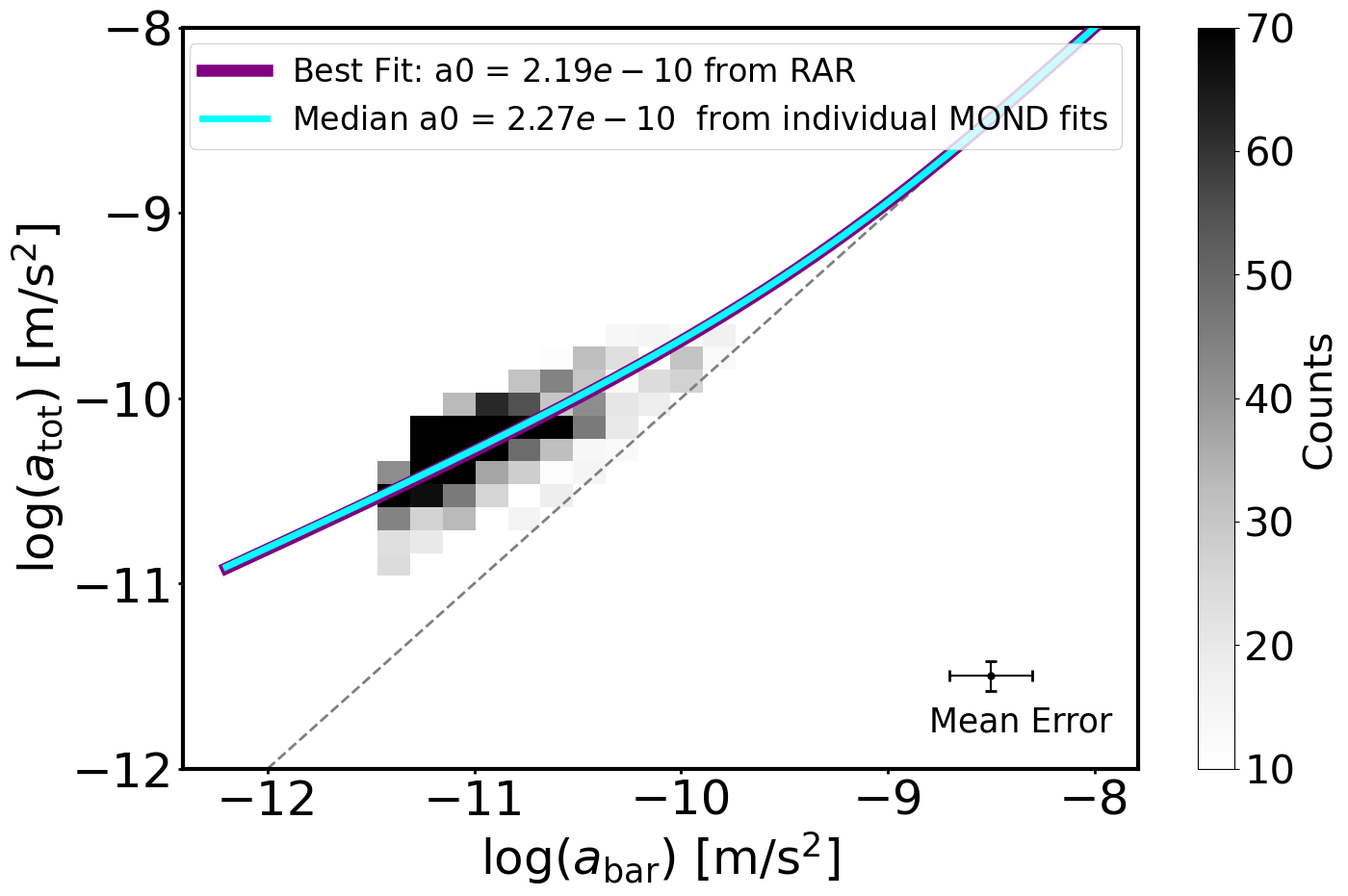}
 \centering
\caption{RAR of our intermediate-$z$ sample, where $a_{\rm bar}$ and $a_{\rm tot}$ are derived from the  RCs modelled in the MOND framework. 
  The cross in the lower right corner indicates the mean $1\sigma$ uncertainties on the data points.
The dotted black line shows the 1:1 relation. The purple curve represents the best-fit RAR, with $a_0|_{z\sim1}  = 2.19^{+0.12}_{-0.10}  \times 10^{-10}~\mathrm{m/s^2}$  (i.e. as in Fig. \ref{rarr}). The cyan curve shows the relation, as inferred using the median $a_0$ value from the MOND fits to the individual data cubes, namely $ a_0|_{z\sim1} = 2.27 \times 10^{-10}~\mathrm{m/s^2}$ .
}
\label{RAR_MOND}
\end{figure}

Finally, for consistency, we re-analyse the RAR from these new fits,  using the results from the MOND model (i), i.e. with $a_0$ free in each galaxy and $\delta\sim1$, following the methodology presented in Sect. \ref{results}. First, we derive $a_{\rm bar}(r)$ and $a_{\rm tot}(r)$ as described in Appendix \ref{methodology} from the individual model RCs. The resulting RAR is shown in  Fig.~\ref{RAR_MOND}. We then fit the resolved RAR tracks of our sample with Eq. \ref{RAR}, leaving $a_0$ as a free parameter (using the same methodology as in Sect. \ref{Res:MHUDF RAR}), and we show the result as the purple curve in Fig.~\ref{RAR_MOND} and as the blue point in Fig.~\ref{comp}, namely:
\begin{eqnarray}
a_0|_{z\sim1} = 2.19^{+0.12}_{-0.10} \times 10^{-10}~\mathrm{m/s^2}
\end{eqnarray}
The recovered $a_0|_{z\sim1}$ value is in fairly good agreement with the values of $a_0|_{z\sim1}$ obtained previously when the data were fitted with DM halos (Eq.~\ref{eq:main:a0}, purple data point in Fig.~\ref{comp}), and is larger than the canonical $z=0$ value (red star in Fig.~\ref{comp}).
In Fig.~\ref{RAR_MOND}, we also plot as the cyan curve Eq.\ref{RAR} using for the characteristic acceleration scale the median value inferred for the sample from the individual MOND fits on the data cubes, i.e. $a_0|_{z\sim1} = 2.27 \times 10^{-10}~\mathrm{m/s^2}$, which close to the $a_0|_{z\sim1}$ inferred from the fit to the RAR.


Following the procedure presented in Sect.~\ref{Res:evol}, we also fit the RAR extracted from the MOND framework using the $z$-dependent acceleration scale (Eq. \ref{RAR} substituting $a_0$ with Eq. \ref{andreea}), and obtain: 
\begin{equation}
a_0(0) = 1.03 ^{+0.05}_{-0.05} \times 10^{-10} ~\mathrm{m/s^2}; \:
a_1 = 1.20^{+0.10}_{-0.10} \times 10^{-10}~\mathrm{m/s^2}.
\label{eq:mond:a1}
\end{equation}

To test whether this result is consistent with the values inferred from individual MOND fits, we also performed a direct linear regression of the best-fit $a_0$ values from each galaxy as a function of $z$, which yields a slope $a_1 = 1.42^{+0.94}_{-0.89}\times 10^{-10}\ \mathrm{m\,s^{-2}},$ and an intercept $a_0(0)= 1.11 ^{+0.39}_{-0.51}\times 10^{-10}\ \mathrm{m\,s^{-2}}$,
in fairly good agreement with the $z$-dependent fit to the RAR presented above (Eq.~\ref{eq:mond:a1}), and with the values obtained in the main text (Eq.~\ref{eq:fits:a1}).  
Together, these results reinforce the presence of a systematic increase of the characteristic acceleration scale with redshift as we found previously when the data were fitted with DM halos (Sect.~\ref{Res:evol} and Appendix~\ref{RARothermodel}).

We end this appendix by noting that these MOND fits underperform compared to the DM-based models (Appendix \ref{RARothermodel}) for over 60\% of the sample
using the  Bayesian model comparison test as in \citetalias{paperI}.

\section{Quantifying the $z$-evolution of the RAR}
\label{appendix:corner}

\begin{figure}
     \includegraphics[width=9cm]{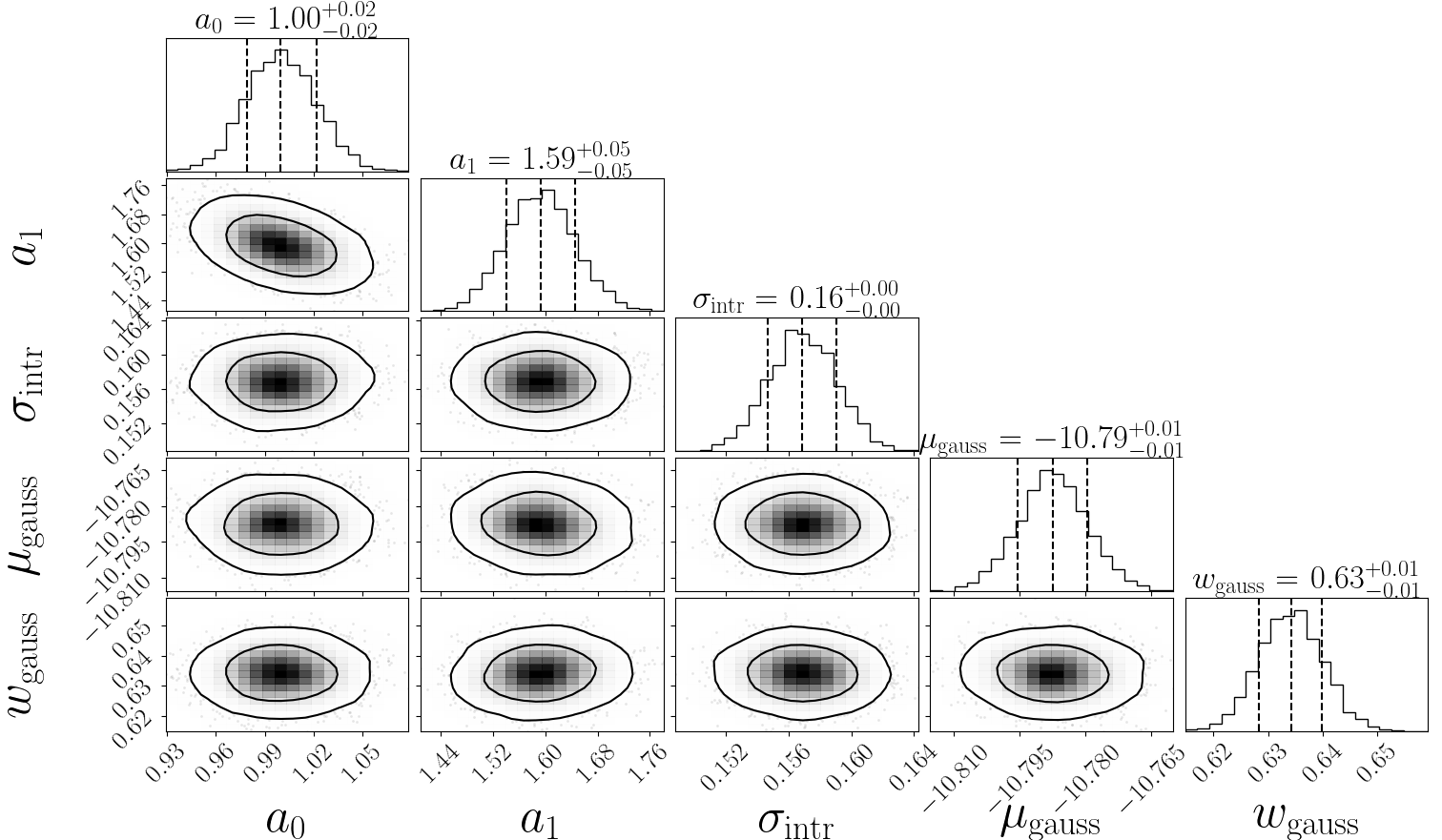}
     \caption{Corner plot showing the posterior distributions of the parameters from the redshift-dependent fit to the RAR for the full sample. The acceleration scale is replaced by $a_0(z) = a_0(0) + a_1 \cdot z$ in the MOND-inspired interpolation function (Eq.~\ref{RAR}).  The Gaussian hyperparameters $\mu_{\rm gauss}$ and $w_{\rm gauss}$ define the mean and width of the Gaussian prior on the true $\log(a_{\rm bar})$ values, as implemented in \texttt{Roxy}, while $\sigma_{\rm intr}$ represents the intrinsic scatter of the RAR. Units of $a_0$ and $a_1$ are $10^{-10}~\mathrm{m/s^2}$, while for $\mu_{\rm gauss}$, $w_{\rm gauss}$ and $\sigma_{\rm intr}$ the units are in dex. 
 Contours indicate the 68\% and 95\% confidence levels.}
    \label{corner}
\end{figure}

To complement the $z$-dependent RAR fit presented in Sect.~\ref{Res:evol}, we show the full posterior distributions of all fitted parameters in Fig.~\ref{corner}. The fit was performed on the whole sample using the MNR method implemented in \texttt{Roxy} \citep{roxy}, by replacing $a_0$ in Eq. \ref{RAR} with the redshift-dependent form defined in Eq. \ref{andreea}. 
In addition to the primary parameters, $a_0(0)$ and $a_1$, the corner plot displays the intrinsic scatter of the relation, $\sigma_{\rm intr}$, and the Gaussian hyperparameters $\mu_{\rm gauss}$ and $w_{\rm gauss}$. Here, $\mu_{\rm gauss}$ and $w_{\rm gauss}$ represent the mean and standard deviation of the Gaussian prior on the true $\log(a_{\rm bar})$ values. The hyperparameters are also marginalised over when inferring $a_0(0)$ and $a_1$.  The 2D contours reveal no significant correlations, confirming that $a_0(0)$ and $a_1$ are well constrained. 

This linear relation,  using the best-fit values of  $a_{0}(0)$ and $a_{1}$ is shown as the purple line in Fig.~\ref{rar_z}.  Note that this global fit  is obtained from a single likelihood analysis of all individual data points and is not a regression through the $a_0$ values obtained from the binned analysis, i.e. the black data points shown in Fig.~\ref{rar_z}. Because data points enter the fit with their full measurement uncertainties, higher-$z$ measurements typically carry less statistical weight, and small offsets between the global relation and the binned estimates are therefore expected.

\end{appendix}

\end{document}